# The dynamics of sex ratio evolution

## From the gene perspective to multilevel selection


Krzysztof Argasinski

Institute of Environmental Sciences, Jagiellonian University

Gronostajowa 7, 30-387 Kraków, Poland

e-mail: krzysztof.argasinski@uj.edu.pl, argas1@wp.pl

Tel. +48604535595

Fax: (+48) 12 664 69 12






**ABSTRACT**


The new dynamical game theoretic model of sex ratio evolution emphasizes the role of males as passive carriers of sex ratio genes. This shows inconsistency between population genetic models of sex ratio evolution and classical strategic models. In this work a novel technique of change of coordinates will be applied to the new model. This will reveal new aspects of the modelled phenomenon which cannot be shown or proven in the original formulation. The underlying goal is to describe the dynamics of selection of particular genes in the entire population, instead of in the same sex subpopulation, as in the previous paper and earlier population genetics approaches. This allows for analytical derivation of the unbiased strategic model from the model with rigorous non-simplified genetics. In effect, an alternative system of replicator equations is derived.  It contains two subsystems: the first describes changes in gene frequencies (this is an alternative unbiased formalization of the Fisher-Dusing argument), whereas the second describes changes in the sex ratios in subpopulations of carriers of genes for each strategy. An intriguing analytical result of this work is that the fitness of a gene depends on the current sex ratio in the subpopulation of its carriers, not on the encoded individual strategy. Thus, the argument of the gene fitness function is not constant but is determined by the trajectory of the sex ratio among carriers of that gene. This aspect of the modelled phenomenon cannot be revealed by the static analysis. Dynamics of the sex ratio among gene carriers is driven by a dynamic "*tug of war*" between female carriers expressing the encoded strategic trait value and random partners of male carriers expressing the average population strategy (a primary sex ratio). This mechanism can be




called "*double-level selection"*. Therefore, gene interest perspective leads to multi-level selection.

## INTRODUCTION

Sex ratio evolution is one of the basic examples of evolutionary mechanisms that are presented in every course on evolutionary biology. The first approach to this problem was presented by German biologist Carl Dusing [2]. Historically, it was the first application of mathematical modeling to evolutionary phenomena. Dusing argued that the fitness of females using different sex ratio strategies can be described by the number of their grandoffspring. A similar approach was applied by Fisher and Shaw and Mohler [3, 4, 5]. This is also an important example in evolutionary game theory, known as a *sex ratio game* [6, 7, 8, 9, 10, 11]. The general prediction of this approach is that the sex ratio of 0.5 is evolutionarily stable. However, there is an alternative approach to the modeling of sex ratio evolution related to population genetics [5, 12, 13, 14]. This approach is focused on tracing the genes encoding sex ratio strategies. Those models predict a stable structure of the population describing gene frequencies among males and females and a sex ratio as the effect of expression of those genes. Therefore, there is a major difference between the strategic phenotypic approach and genetic modeling [15, 16, 17]. The phenotypic approach describes the mean female strategy of 0.5 as evolutionarily stable, while genetic models show that the composition of the male population can also matter. To analyze this problem, in our previous paper [1], a new model of sex ratio evolution was developed. The new approach is an attempt to combine the genetic and phenotypic approach and to overcome the limitations of both of them. The goal was to solve the problem of different predictions and to obtain a coherent picture of the modeled phenomenon.



The new model focuses on the global dynamics of the system, and its structure resembles the genetic approach [5, 12, 13, 14]. Whereas the classical Dusing-Fisher-Shaw-Mohler (DFSM) model is focused on the reproductive success of individual strategies carried by female strategic agents (as in Dusing's paper, see [2], or the sex ratio game) or some undescribed group of "parents" (as in [3, 4], more on this topic in section 4.2). For a closer understanding of the relations between the classical and the new approach, the selection of individual strategies resulting from global dynamics must be analyzed, which is the subject of this paper.

In this paper a novel technique of change of coordinates will be applied to the model from [1]. This will reveal new aspects of the modelled phenomenon which cannot be shown or proven in the original formulation. Similarly the results from [1] will be hard to show in the new coordinates, thus the two papers complement each other. The underlying goal is to describe the dynamics of selection of particular genes in the entire population, instead of in the same sex subpopulation as in the previous paper and earlier population genetics approaches. In effect, an unbiased strategic model will be analytically derived from the non-simplified rigorous genetic model.

Thus, the classical strategic approach analyzes the reproductive success of a female, while the genetic approach traces gene frequencies in the population. Therefore, what happens when we combine both perspectives and assume that the gene is the strategic agent?

**METHODS**

Now we shall recall the structure of the new model (see Table 1 for the list of symbols). Section 1 can be skipped by readers familiar with paper [1].

### 1.1 Summary of basic formal details of the new model

There are $u$ individual strategies described by $P_i \in [0,1]$, the proportion of male offspring of a



female playing strategy $P_i$. There are $x_i$ females and $y_i$ male carriers of the strategy $P_i$ in the population. Therefore, the population consists of $x = \sum_i x_i$ females and $y = \sum_i y_i$ males. Thus, $f = [f_1,...,f_u]$ is the vector of frequencies of strategies of the female subpopulation, and $m = [m_1,...,m_u]$ is an analogous vector for the male subpopulation, where $f_i = \dfrac{x_i}{x}$ and $m_i = \dfrac{y_i}{y}$.

$P = \dfrac{y}{y+x}$ is the fraction of males in the population (the secondary sex ratio), and $\sum_j f_j P_j$ is the mean female strategy (the primary sex ratio). Assume that each female produces $k$ offspring according to haploid inheritance. However, males are gene carriers too, and transfer those genes to their offspring with the probability 0.5. The influence of males can be described by the *fitness exchange effect* (i.e. daughters of male carriers contribute to the fitness of female carriers and sons of female carriers contribute to the fitness of male carriers). In [1] it was shown that

$W_{mm} = 0.5 \left( \sum_j f_j P_j \right) \dfrac{xk}{y}$ is the expected number of male offspring, and $W_{mf} = 0.5 \left( \sum_j f_j \left(1 - P_j\right) \right) \dfrac{xk}{y}$ is the expected number of female offspring of the male individual. Analogously,

$W_{fm} = 0.5(1 - P_i)k$ is the expected number of male offspring, and $W_{ff} = 0.5 P_i k$ is the expected number of female offspring of the female individual playing the strategy $P_i$. Therefore, the following equations were obtained:

$$W_m(P_i, P, f, m) = W_{mm} + \frac{x_i}{y_i} W_{fm} = k \frac{1-P}{2P} \left( \sum_j f_j P_j + \frac{f_i}{m_i} P_i \right), \qquad (1)$$

– payoff function of the males carrying the strategy $P_i$,

$$W_f\left(P_i, P, f, m\right) = W_{ff} + \frac{y_i}{x_i} W_{mf} = \frac{k}{2} \left( \left(1 - P_i\right) + \frac{m_i}{f_i} \left(1 - \sum_j f_j P_j\right) \right), \qquad (2)$$

– payoff function of the females playing the strategy $P_i$.



Now we have all elements needed to formulate multipopulation replicator dynamics (see appendix A). In [1], this took the following form:

$$\dot{f}_i = f_i\big(W_f(P_i, P, f, m) - \overline{W}_f(P, f, m)\big) \quad \text{for} \quad i = (1, \ldots, u-1),$$
$$\dot{m}_i = m_i\big(W_m(P_i, P, f, m) - \overline{W}_m(P, f, m)\big) \quad \text{for} \quad i = (1, \ldots, u-1),$$
$$\dot{P} = P\big(\overline{W}_m(P, f, m) - \overline{W}(P, f, m)\big),$$

where $\overline{W}_m(P, f, m) = \sum_i m_i W_m(P_i, P, f, m)$, $\overline{W}_f(P, f, m) = \sum_i f_i W_f(P_i, P, f, m)$,

$\overline{W}(P, f, m) = P\overline{W}_m(P, f, m) + (1-P)\overline{W}_f(P, f, m)$ are the respective average payoff functions of the male, female and the whole population. This leads to the following system of equations:

$$\dot{f}_i = k\left(\frac{f_i}{2}(1 - P_i) + \left(\frac{m_i}{2} - f_i\right)\left(1 - \sum_j f_j P_j\right)\right) \quad \text{for} \quad i = (1, \ldots, u-1),$$

$$\dot{m}_i = \frac{k}{2}\left(\frac{1-P}{P}\right)\left(f_i P_i - m_i \sum_j f_j P_j\right) \quad \text{for} \quad i = (1, \ldots, u-1),$$

$$\dot{P} = k(1-P)\left(\sum_j f_j P_j - P\right).$$

It was shown that, for biological reasons, we can limit the analysis of the model to values of primary and secondary sex ratios over the interval $(0,1)$.

## 1.2 Summary of predictions of the new model

An analysis of the behavior of this model shows that two phases of convergence can be distinguished. The first, rapid phase occurs when the secondary sex ratio $P$ converges to the current value of the primary sex ratio $\sum_j f_j P_j$, and the male subpopulation converges to the state termed the male subpopulation equilibrium (MSE), described by the condition $f_i P_i = m_i \sum_j f_j P_j$. During the second phase of convergence, the primary sex ratio converges to the value 0.5, and the value of the secondary sex ratio follows these changes to maintain equality. In addition, the state



of the male subpopulation changes to maintain the MSE.

**RESULTS**

## 2. Reformulation of the model.

In the previous paper [1], a change in the coordinates (described in appendix A) was applied to the numerical solutions obtained to calculate the frequencies of all types of individuals (see Fig. 3c in [1] and section 3.2 there) and gene frequencies (see Fig. 6 in [1] and section 4 there). However, this method can be applied not only to numerical solutions, but also directly to replicator equations. In this way, we can reformulate the new model to focus on changes in gene frequencies. We have $Pm_i$ male carriers and $(1-P)f_i$ female carriers of a strategy $P_i$ in the whole population. Thus, the frequency of carriers of a gene which encodes this strategy is equal to:

$$G_i = Pm_i + (1-P)f_i. \tag{3}$$

The state of the population can be described by the vector $G = [G_1,...,G_u] \in \Delta^u$, where $\sum_j G_j = 1$. In this description, there is no information about the sex of the carriers of these genes. We can fill this gap by adding information about the sex ratio in the subpopulation of the carriers for every gene:

$$M_i = \frac{Pm_i}{Pm_i + (1-P)f_i} = \frac{Pm_i}{G_i} \qquad \text{-proportion of males among carriers of } P_i,$$

$$\tag{4}$$

$$F_i = 1 - M_i = \frac{(1-P)f_i}{Pm_i + (1-P)f_i} = \frac{(1-P)f_i}{G_i} \qquad \text{-proportion of females among carriers of } P_i.$$

Then, $M = [M_1,...,M_u]$ is the vector of subpopulation sex ratios. Therefore, this structure can be treated as a division of the entire population into $u$ subgroups with one-dimensional subpopulation states. Then, according to the general notation from appendix A, $\sigma^i = M_i$ and



$\gamma_j = G_j$ (see also [18]), the structure of the space of population states will take the form

presented in Fig. 1.

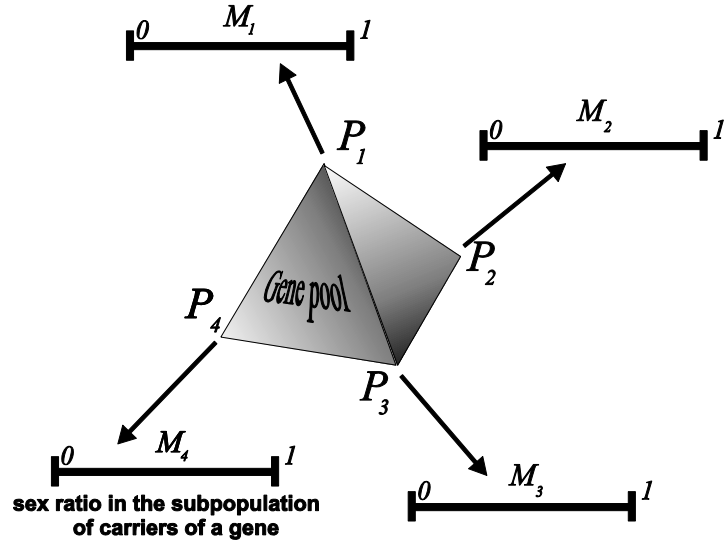

**Fig. 1 Scheme of a space of population states in the new formulation of the model. In this case, it is a product of a simplex of gene frequencies and $u$ one-dimensional simplexes that describe sex ratios in the subpopulations of carriers for each strategy.**

Note that in the previous formulation of the model, the space of population states was the

product of two $u-1$ dimensional simplexes of the male and female subpopulation and a one-

dimensional simplex of the proportion between these subpopulations (a secondary sex ratio); in

general, the dimension of the whole space was $2u-1$. In the new formulation, this space consists

of one $u-1$ dimensional simplex of gene frequencies and $u$ one-dimensional simplexes of

subpopulation sex ratios, and the dimension of the whole space of population states is also

$2u-1$. Therefore, the dimension of the space of population states is invariant in response to the

change of coordinates, which is consistent with the fact that we have a different parameterization

of the same phase space. We can describe important population parameters in the new

coordinates for parameters such as the mean female subpopulation strategy $\overline{P}_{pr}$, i.e., the primary

sex ratio and secondary sex ratio (among adult individuals) $P$:



$$\overline{P}_{pr} = \sum_j f_j P_j = \frac{1}{1-P} \sum_j (1-M_j) G_j P_j \quad \text{and} \quad P = \sum_j G_j M_j .$$

The average fitness functions from the previous paper (recalled in section 1.1) were:

$$\overline{W}_m(P,f,m) = k \frac{1-P}{P} \sum_j f_j P_j \quad \text{– mean fitness of the male subpopulation,}$$

$$\overline{W}_f(P,f,m) = k\left(1 - \sum_j f_j P_j\right) \quad \text{– mean fitness of the female subpopulation,}$$

$$\overline{W}(P,f,m) = k(1-P) \quad \text{– mean fitness of the whole population.}$$

Then, we can derive the mean payoff to the carrier of a gene for strategy $P_i$ (for a full derivation see appendix B):

$$W_g(P_i,P,f,m) = W_g(P_i,G,M) = M_i W_m(P_i,P,f,m) + (1-M_i)W_f(P_i,P,f,m) ,$$

which takes the form:

$$W_g(P_i,G,M) = \frac{k}{2}\left(\frac{(1-P)}{P} M_i + \left(1-M_i\right)\right) = \frac{k}{2}\left(\Gamma M_i + \left(1-M_i\right)\right), \tag{5}$$

where $\Gamma = \frac{1-P}{P}$ is the number of females per single male individual. For the new coordinates we obtain the following replicator equations (for a detailed derivation, see appendix C):

$$\dot{G}_i = G_i\left(W_g(P_i,P,f,m) - \overline{W}(P,f,m)\right) \quad \text{-dynamics of gene frequencies,}$$

$$\dot{M}_i = M_i\left(W_m(P_i,P,f,m) - W_g(P_i,P,f,m)\right) \quad \text{-dynamics of sex ratios in carriers subpopulations,}$$

which take the form:

$$\dot{G}_i = G_i k\left(\frac{1}{2} - P\right)\left(\frac{M_i}{P} - 1\right) \quad \text{for } i=(1,...,u-1), \tag{6}$$

$$\dot{M}_i = \frac{k}{2}\left(M_i\left(\frac{1-P}{P}\right)\left(\overline{P}_{pr} - M_i\right) + \left(1-M_i\right)\left(P_i - M_i\right)\right) \quad \text{for } i=(1,...,u). \tag{7}$$

## 3. Behavior of trajectories of replicator equations

## 3.1 Trajectories of gene frequencies



Here, we will examine the dynamics of gene frequencies. The product $\left(\frac{1}{2} - P\right)\left(\frac{M_i}{P} - 1\right)$ is responsible for the sign of the right side of equation (6). When both coefficients are negative or positive, then their product is positive (the frequency of gene $P_i$ increases), and when they have opposite signs, then their product will be negative (the frequency of gene $P_i$ decreases). The zero points of these coefficients, $P = \frac{1}{2}$ and $P = M_i$, are stationary points of equation (6). Therefore, the dynamics of the gene frequencies can be described in the following way:

$G_i$ increases when $P < \frac{1}{2}$ and $P < M_i$ or $P > \frac{1}{2}$ and $P > M_i$,

$G_i$ decreases when $M_i < P < \frac{1}{2}$ or $M_i > P > \frac{1}{2}$, $\hspace{3cm}$ (8)

$G_i$ is constant when $G_i = 0$ or $M_i = P$ or $P = \frac{1}{2}$.

Recall that $P = \sum_j G_j M_j$, which means that the secondary sex ratio is equal to the average sex ratio in the carrier subpopulation over the entire population. Therefore, the frequency $G_i$ decreases when the sex ratio in the carrier subpopulation $M_i$ is shifted farther from 0.5 than the mean sex ratio in the carrier subpopulations for all strategies $P$. In the opposite case, $G_i$ will increase. This mechanism is illustrated in Fig. 2a. Therefore, the frequency of a gene that encodes the strategy 0.5 increases when the sex ratio in a subpopulation of its carriers is closer to 0.5 than the current value of the secondary sex ratio; this frequency decreases in the opposite case. A situation in which the secondary sex ratio is equal to 0.5 is the stationary state of the dynamics of gene frequencies (6). Therefore, this mechanism described by (8) is independent of individual strategies $P_i$, but its dynamics are dependent on the trajectories of the sex ratios in the subpopulations of carriers of the strategies described by $M_i$.



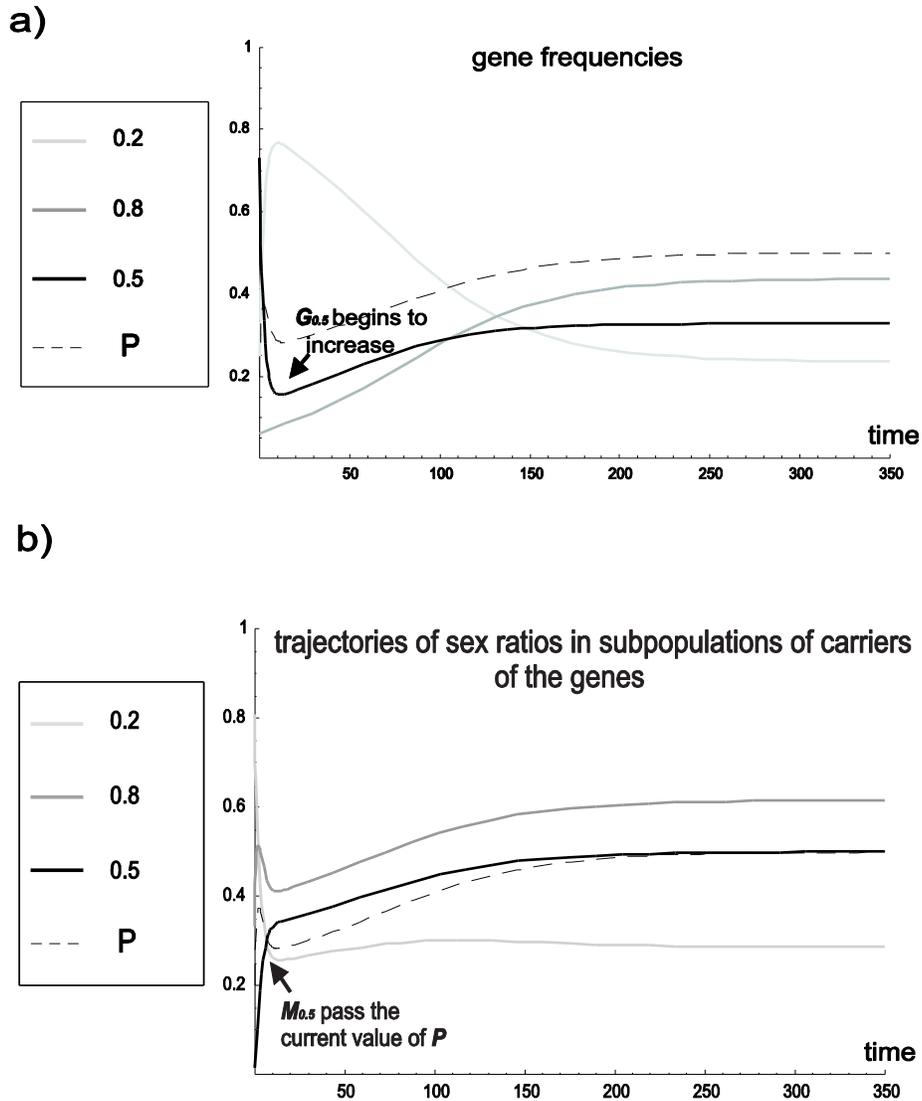

**Fig. 2 Trajectories of a population of individuals with strategies for sex ratios of *0.2*, *0.5* and *0.8* for initial conditions** $M_{0.2}$=0.81, $M_{0.5}$=0.01, $M_{0.8}$=0.33 and $G_{0.2}$=0.21, $G_{0.5}$=0.73, $G_{0.8}$=0.06. **Panel *a*) shows the trajectories of gene frequencies** $G_i$, **Therefore,** $G_i$ **increases when** $P < \dfrac{1}{2}$ **and** $P < M_i$ **or** $P > \dfrac{1}{2}$ **and** $P > M_i$ **and decreases when** $P < \dfrac{1}{2}$ **and** $P > M_i$ **or** $P > \dfrac{1}{2}$ **and** $P < M_i$. **This mechanism is clearly shown in the trajectories of strategy *0.5*. The trajectory** $G_{0.5}$ **switches from a decrease to an increase when trajectory of** $M_{0.5}$ **passes the trajectory of** $P$ **(see panel *b*). Panel *b*) shows the respective changes of sex ratios in carrier subpopulations** $M_i$. **Note that sex ratios in carrier subpopulations rapidly converge to the values determined by the *MSE* phenomenon, and after that, they follow the changes of the primary sex ratio** $\overline{P}_{pr}$ **that slowly converges to *0.5*. The sex ratio among carriers of male biased strategies change due to the dynamics of the primary sex ratio while among female biased strategies, it converges to the neighbourhood of the value encoded by the gene.**



Note that parameter $M_i$ also affects the secondary sex ratio $P = \sum_j G_j M_j$, modifying the values of $G_j$. However, sex ratios in carrier subpopulations $M_j$ are determined by mechanisms acting at the level of carrier subpopulations that are described in the next section.

## 3.2 Trajectories of sex ratios in subpopulations of carriers

The dynamics of sex ratios in the carrier subpopulations are more sophisticated. The right side of equation (7) contains two coefficients: $\left(\overline{P}_{pr} - M_i\right)$ and $\left(P_i - M_i\right)$, weighted by current values of $M_i \Gamma$ and $\left(1 - M_i\right)$. These coefficients are responsible for the direction of convergence. The coefficient $\left(\overline{P}_{pr} - M_i\right)$ induces attraction of $M_i$ to $\overline{P}_{pr}$, and the coefficient $\left(P_i - M_i\right)$ causes attraction of $M_i$ to $P_i$. This is, in a sense, a *tug of war* between female partners of the male carriers (representing average strategy $\overline{P}_{pr}$) and female carriers of the same gene (representing encoded strategy $P_i$). As we can see in Fig. 2b, the shape of the trajectory of a 0.8 sex ratio strategy that produces mostly sons is almost parallel to the trajectory of parameter $P$, which is equal to $\overline{P}_{pr}$ in the slow phase of convergence (see [1]). On the other hand, the trajectory of a 0.2 sex ratio strategy that produces more daughters is closer to the constant function 0.2 than to the trajectory of $P$. Thus, the $M_i$ value of the strategies producing (and in effect carried by) mostly males resemble trajectories of the primary sex ratio, while female biased strategies have $M_i$ almost constant and equal to $P_i$. This interesting aspect would be hard to show by static analysis. Below, we will characterize equilibrium in this "tug of war".

**Lemma 1**



a) For every set of values of $P$, $\overline{P}_{pr} \in (0,1)$ and $P_i \in (0,1]$, dynamics (7) has the unique

stable conditional equilibrium $\overline{\overline{M}}_i$ that is contained in the interval limited by the values of

$\overline{P}_{pr}$ and $P_i$.

b) For the strategy $P_i = 0$, there is one stationary point, $\overline{\overline{M}}_i = 0$, which is stable when

unique. However, when $\overline{P}_{pr} > \dfrac{1}{\Gamma}$ and $P < \frac{1}{2}$, the rest point $\overline{\overline{M}}_i = 0$ becomes unstable, and

there exists a second stationary point $\overline{\overline{M}}_i = \dfrac{\overline{P}_{pr}\Gamma - 1}{\Gamma - 1}$.

For a proof, see Appendix D.

Lemma 1 indicates that, at every moment, there exists some attracting point for $M_i$ lying between

the current value of the primary sex ratio $\overline{P}_{pr}$ (which also changes in time) and the value of

individual strategy $P_i$. By this dynamic equilibrium, the expression of individual strategies

determines the parameter $M_i$. The only exception is strategy $P_i = 0$ (production of female

offspring only) for which the second stationary state may exist during the rapid phase

of convergence. It was impossible to analytically derive the stable sex ratio in the carrier

subpopulations, in the general case. This is possible only when the population is in the MSE state

and will be presented in a subsequent paper devoted to the MSE. According to Lemma 1, we can

numerically approximate this value because it is unique in these biologically significant cases.

## DISCUSSION

## 4.1. The mechanism of "double-level" selection

Here, we will summarize the results we have obtained. The first intriguing analytical result of the



reformulated model is that the fitness function of a gene (5) is independent of the individual strategy it encodes. Proliferation of a given gene depends on the current sex ratio in the subpopulation of its carriers $M_i$. Note that the fitness function (5) is a good mathematical description of Fisher's idea, which is related to the reproductive value of carriers with different sexes according to the deviation of the secondary sex ratio *P*. It suggests that males are reproductively more efficient when they are in the minority (*P<1/2*), because each male can mate with several females (Γ>1). On the other hand, females are more efficient when they are in the minority (*P>1/2*), because each female will be expected to produce offspring, and there are not enough mates for all males (Γ<1). Therefore, parameter $M_i$ describes the proportion of carriers with the more reproductively efficient sex among all carriers of a gene. This fitness function explicitly considers male carriers from the mother's generation of unexpressed sex ratio genes. Function (5) can be transformed in the following way (recall that $y_i$ is the number of male carriers, and $x_i$ is the number of female carriers, of the strategy $P_i$):

$$W_g = \frac{k}{2}\left(\Gamma M_i + \left(1 - M_i\right)\right) = \frac{k}{2}\left(\frac{y_i}{x_i + y_i}\Gamma + \frac{x_i}{x_i + y_i}\right) = \frac{1}{x_i + y_i}\left(y_i\Gamma\frac{k}{2} + x_i\frac{k}{2}\right).$$

This is the per capita normalized sum (averaged over the carriers subpopulation) of the offspring produced by female partners of male carriers described by $y_i\Gamma\frac{k}{2}$ and offspring of female carriers described by $x_i\frac{k}{2}$ (where $\frac{k}{2}$ is the number of offspring of a single female multiplied by the probability of gene transfer from the focal parent). This is an explanation of the importance of male carriers of the unexpressed sex ratio genes, or rather their female partners. Their role is important, because each male carrier may have Γ partners, and the activity of their partners is an important component of gene fitness. Surprisingly, this function is independent of the value of a



given strategy, $P_i$, encoded by the carried gene. It depends only on $\Gamma$ and $M_i$. The phenomenon can be termed *double level selection*. The fitness of a gene that encodes an individual strategy is determined in some way by the current sex ratio in its carrier subpopulation and the secondary sex ratio in the population as a whole. Values of both parameters may be perturbed. However, the stable carrier subpopulation sex ratio should be determined in some way by the value of the encoded strategy (Fig. 3). This is a newly discovered mechanism. In general, the mechanism of double level selection can be regarded as an example of *multi-level selection*, which is the concept presented by [19, 20, 21, 22, 23]. The classical approach to the modeling of sex ratio evolution treats this phenomenon as *single level* selection, which means that the fitness is unambiguously determined by the values of individual strategy $P_i$ and a population state described by the secondary sex ratio (Fig. 3). In the next subsection, a higher level of this process will be considered.

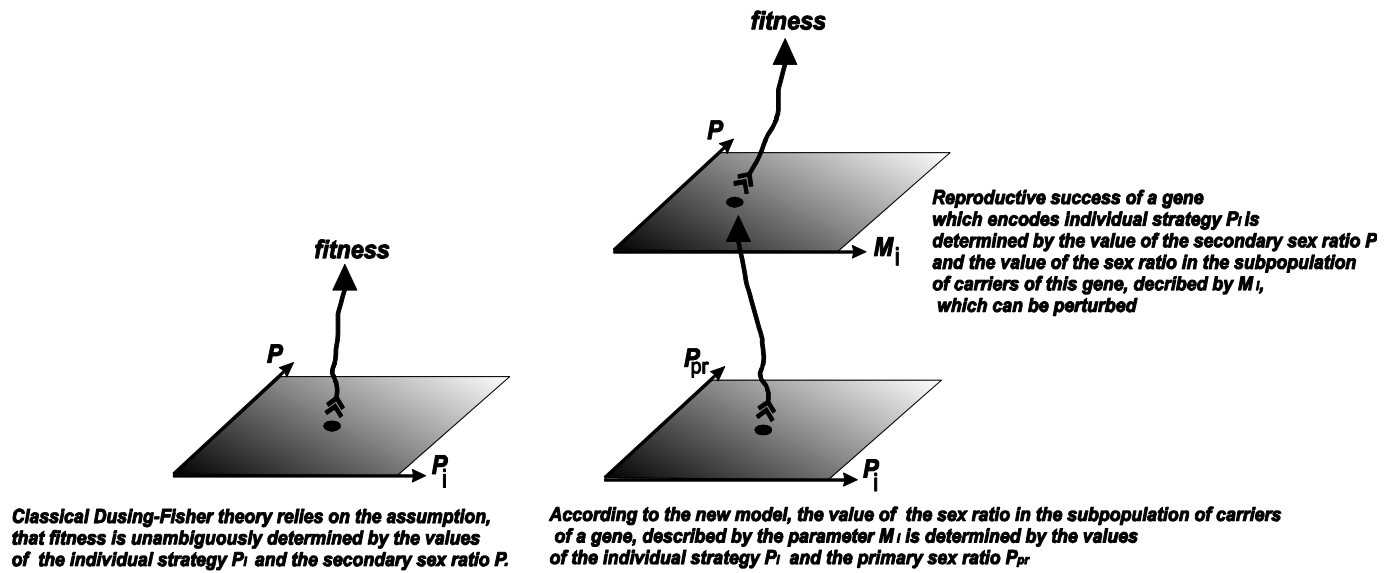

Classical Dusing-Fisher theory relies on the assumption, that fitness is unambiguously determined by the values of the individual strategy $P_i$ and the secondary sex ratio P.

Reproductive success of a gene which encodes individual strategy $P_i$ is determined by the value of the secondary sex ratio P and the value of the sex ratio in the subpopulation of carriers of this gene, decribed by $M_i$, which can be perturbed

According to the new model, the value of the sex ratio in the subpopulation of carriers of a gene, described by the parameter $M_i$ is determined by the values of the individual strategy $P_i$ and the primary sex ratio $P_{pr}$

**Fig. 3. A comparison of "single level" selection and "double level" selection.**



## 4.2 Dynamics of gene frequencies

The mechanism realized by gene frequency replicator equations (6), described by the rules (8) increases the frequency of a gene for which the value of a parameter $M_i$ is greater/smaller than the secondary sex ratio $P$ (which is equal to the average $M$ in the population) when $P$ is smaller/greater than 0.5. Thus, it is profitable for the gene to be carried by that sex which is currently in the minority. There is an interesting relationship between the mechanism described by (8) and the replicator dynamics paradigm. In standard replicator equations, frequencies of strategies change according to the sign of the deviation of their fitness from average fitness (minus - decrease, plus - increase). If fitness depends linearly on a particular trait, then selection works according to deviations from the average trait value. Note that the payoff function (5) is linear with respect to the parameter (trait) $M_i$, and the secondary sex ratio $P$ is an average $M_i$ over the population. The difference between the mechanism in rules (8) and standard replicator dynamics is that parameter $M_i$ is not a description of a fixed individual strategy but of the current state of a subgroup of individuals (the subpopulation of carriers of strategy $P_i$). Dusing classically argued that female producing offspring of the sex that is currently in the minority will have more grand-offspring. This argument states that there are differences in fitness among females with different strategies, which is considered a proof of the existence of selection on individual strategies. However, our new model shows that a mechanism based on different reproductive values is independent of individual strategies $P_i$, and it affects the primary sex ratio $\overline{P}_{pr}$ and the secondary sex ratio $P$ (which is equal to the average sex ratio in the carrier subpopulation) by changing only gene frequencies $G_j$. In [3], the following statement can be



found:

*"...it would follow that those parents, the innate tendencies of which caused them to produce males in excess, would for the same expenditure, produce a greater amount of reproductive value; and in consequence would be the progenitors of a larger fraction of future generations..."*.

Therefore, Fisher in his original reasoning considered a group of individuals that adjusts the sex ratio among its members due to genetic mechanisms. However, the mechanisms for this adjustment were not explicitly explained. The perspective of a group adjusting the sex ratio among its members is also assumed by [4]. However, they also presented only a conjecture that the sex ratio is completely heritable within the group, without an explanation of how it is realized. Therefore, there is a difference between Fisher's reasoning that operated on the level of the subpopulation of all carriers of a gene and Dusing's approach related to the level of female individuals. The female perspective is not sufficient, especially for male-biased strategies, which will produce more male than female carriers. This means that the Fisherian argument about the different reproductive values of males and females is an important part of understanding sex ratio self-regulation. However, it is not enough for a full mechanistic explanation of this process. Therefore, we should investigate how the expression of individual strategies determines the sex ratio in the carrier subpopulation $M_i$. This will allow us to overcome the limitations of Dusing's reasoning, which considers only female reproductive success and disregards the role of male gene carriers from the same generation.

## 4.3 Dynamics of sex ratios in carrier subpopulations: the "tug of war" mechanism

The sex ratio in carrier subpopulations is the effect of intrinsic dynamics that can be compared to a "tug of war" between $P_i$ and $\overline{P}_{pr}$. It was proved in Lemma 1 that for every population state



there exists a single unique attractor of $M_i$ dynamics contained in the interval that is limited by values of $\overline{P}_{pr}$ and $P_i$. Let us describe the "tug of war" metaphor in a more formal way. The right-hand side of replicator equation (7) is proportional to

$$M_i \left( \frac{1-P}{P} \right) \left( \overline{P}_{pr} - M_i \right) + \left( 1 - M_i \right) \left( P_i - M_i \right).$$

The factor $M_i \left( \dfrac{1-P}{P} \right)$ that is the weight of $\left( \overline{P}_{pr} - M_i \right)$ can be written as $\dfrac{y_i \Gamma}{x_i + y_i}$ and the proportion $\left( 1 - M_i \right)$ that is the weight of $\left( P_i - M_i \right)$ equals $\dfrac{x_i}{x_i + y_i}$. Thus the right side of this equation is proportional to

$$\frac{1}{x_i + y_i} \left( y_i \Gamma \left( \overline{P}_{pr} - M_i \right) + x_i \left( P_i - M_i \right) \right).$$

Since $\Gamma$ is the number of females per single male, then $y_i \Gamma$ is also the number of female partners of male carriers of gene encoding the strategy $P_i$. These females "pull the rope" toward the value of $\overline{P}_{pr}$. On the other side, a team of $x_i$ female carriers of this gene "pulls the rope" toward the value $P_i$. It is evident here that the expression of strategies of parental individuals determines the fate of their descendants, by the setting of the sex ratio among them.

## 4.5 An unresolved problem: the role of the male subpopulation equilibrium

Recall that, during the slow phase of the sex ratio dynamics, $\overline{P}_{pr} = P$. Note that, if in rules (8) we substitute $\overline{P}_{pr}$ instead of $P$ and $f_i$ instead of $M_i$ we obtain the following rules:

$f_i$ increases when $\sum_j f_j P_j < \frac{1}{2}$ and $\sum_j f_j P_j < P_i$ or $\sum_j f_j P_j > \frac{1}{2}$ and $\sum_j f_j P_j > P_i$,



$f_i$ decreases when $P_i < \sum_j f_j P_j < \frac{1}{2}$ or $P_i > \sum_j f_j P_j > \frac{1}{2}$,

$f_i$ is stable when: $f_i = 0$ or $f_i = 1$ or $P_i = \sum_j f_j P_j$.

These describe the changes of a female subpopulation state when the MSE condition is satisfied (Lemma 1 from [1]). This leads to the problem of the role of the MSE phenomenon, which is responsible for the rapid phase of convergence and the dynamics of sex ratios in the carrier subpopulations. The first idea that comes to mind to explain this phenomenon is that the male subpopulation equilibrium is equivalent to some stable sex ratio in the carrier subpopulation (the equilibrium of the "tug of war" mechanism), which is conditional on current values of $P$, $\overline{P}_{pr}$ and $P_i$. The rapid phase will then be equivalent to convergence to this stable value. When the subpopulation reaches a stable sex ratio, then it simply follows changes of the primary (and in effect the secondary) sex ratio, which are equivalent to the slow phase of convergence. Unfortunately, this idea is false. As shown in [1], when the MSE conditions are satisfied for all strategies, then all males in the population have the same fitness. If we assume that carrier subpopulations are in their stable states, then for all strategies females will have fitness equal to males. So, when all males have equal fitness, and all females have fitness equal to males, then all individuals in the population have equal fitness. In this case, the population would be in a global stationary state, which is not true. The nature and role of the male subpopulation equilibrium are the subjects of a subsequent paper.

Acknowledgments

I want to thank my supervisor Jan Kozłowski for inspiration and support during my PhD studies. I also want to thank Mark Broom, Jozsef Garay, Josef Hofbauer, Vlastimil Krivan, John McNamara, Jacek Miękisz, Steven Orzack, Ido Pen, Jacek Radwan and Franjo Weissing for



discussions, helpful suggestions and hospitality during my consultation visits performed during the realization of the project. In addition, I want to thank Bob Planque and anonymous Reviewer for valuable suggestions and fast turnover time.

Table 1: List of important symbols:

| **classical theory**: |
|---|
| $P$ - secondary sex ratio |
| $P_{ind}$ - individual strategy interpreted as the mean sex ratio in the brood of a single female, which is the carrier of this strategy ( $P$ with index denotes the individual strategy) |
| $N$- population size |
| $k$ - mean brood size of a single female |
| $\overline{W}_f(P,f,m) = \sum_i f_i W_f(P_i,P,f,m)$ - mean fitness function of the female subpopulation |
| $\overline{W}(P,f,m) = P\overline{W}_m(P,f,m) + (1-P)\overline{W}_f(P,f,m)$ - mean fitness function of the whole population |
| $W(P_{ind},P)$ - classical Dusing-Fisher-Shaw-Mohler fitness function |
| **new model**: |
| $y$ - number of males |
| $x$ - number of females |
| $N = y + x$ - population size |
| $u$ - number of individual strategies |
| $f_i = \dfrac{x_i}{x}$ - frequency of females with strategy $P_i$ |
| $m_i = \dfrac{y_i}{y}$ frequency of males with strategy $P_i$ |
| $f = [f_1,...,f_u]$ -state vector of the female subpopulation |
| $m = [m_1,...,m_u]$ -state vector of the male subpopulation |
| $G = [G_1,...,G_u]$ - state vector of the gene pool |
| $G_i = Pm_i + (1-P)f_i$ - frequency of a gene which encodes the strategy $P_i$ |
| $M_i = \dfrac{Pm_i}{Pm_i + (1-P)f_i}$ fraction of males in the subpopulation of carriers of the strategy $P_i$ |
| $P = \dfrac{y}{y+x}$ - frequency of males in the population |
| $\Gamma = \dfrac{x}{y} = \dfrac{1-P}{P}$ - number of females per single male individual |



$\overline{P}_{pr} = \sum_j f_j P_j$  -primary sex ratio (mean strategy in the female subpopulation)

$W_m(P_i, P, f, m)$  - males' payoff function

$W_f(P_i, P, f, m)$  - females' payoff function

$W_g(P_i, G, M)$ - fitness function of a gene which encodes strategy $P_i$

$\overline{W}_m(P, f, m) = \sum_i m_i W_m(P_i, P, f, m)$   - mean fitness function of the male subpopulation

## Appendix A

## Change of coordinates in the space of population states.

Assume that we want to break down an entire population into $z$ subgroups. Define

$d^i = [d_1^i, ..., d_{u_i}^i]$ as a vector of indices of strategies exhibited by individuals from the $i$-th

subgroup ( $d_j^i \in \{1, ..., u\}$ , $u_i$ the number of strategies in the $i$-th subgroup). For example, the

notation $d^2 = [1, 3, 5]$ means that in the second subgroup, there are individuals with strategies $1, 3$

and $5$. Every strategy should belong to a single unique subgroup (and cannot belong to two).

Then, according to [18] using the following change of coordinates:

$$\sigma^i = [\sigma_1^i, ..., \sigma_{u_i}^i] = \left[ \frac{\sigma_{d_1^i}}{\sum_{j=1}^{u_i} \sigma_{d_j^i}}, ..., \frac{\sigma_{d_{u_i}^i}}{\sum_{j=1}^{u_i} \sigma_{d_j^i}} \right] \quad \text{for} \quad i = 1, ..., z,  \qquad \text{(a1)}$$

we obtain the distribution of relative frequencies of strategies in the $i$-th subpopulation. The

distribution of proportions between subpopulations has the form:

$$\gamma = [\gamma_1, ..., \gamma_z] = \left[ \sum_{i=1}^{u_1} \sigma_{d_i^1}, ..., \sum_{i=1}^{u_z} \sigma_{d_i^z} \right],  \qquad \text{(a2)}$$

where $\gamma_i$ is the proportion of the $i$-th subpopulation. Every decomposition into subpopulations

can be reduced again to a single population model by the opposite change of coordinates

$\sigma(\gamma, \sigma^1, ..., \sigma^z)$ where:



$$\sigma_{d_j^i} = \gamma_i \sigma_j^i. \tag{a3}$$

Note that we can break down an entire population into $z$ subpopulations. When we apply the above transformations to replicator equations, we obtain a set of equations that describes the dynamics inside subpopulations (intraspecific dynamics, see [18]), which has the form:

$\dot{\sigma}_j^i = \sigma_j^i \left[ W_j^i - \overline{W}^i \right]$, where $W_j$ is the fitness of the $j$-th strategy in the $i$-th subpopulation and $\overline{W}^i$ is the mean fitness in the $i$-th subpopulation, and a system that describes changes of relative sizes among subpopulations (interspecific dynamics) is:

$\dot{\gamma}_s = \gamma_s \left[ \overline{W}^s - \overline{W} \right]$, where $\overline{W}$ is the mean fitness in the whole population.

When the set of strategies in each subpopulation is characterized by a vector of indices $d^i$, then the system of replicator equations will be:

$$\dot{\sigma}_j^i(t) = \sigma_j^i(t) \left[ W(P_{d_j^i}, \sigma(\gamma(t), \sigma^1(t), ..., \sigma^z(t))) - \overline{W}^i(\sigma(\gamma(t), \sigma^1(t), ..., \sigma^z(t))) \right] \quad \text{for}$$
$$j = 1, ..., u_i - 1 \quad \text{and} \quad i = 1, ..., z, \tag{a4}$$

$$\dot{\gamma}_s(t) = \gamma_s(t) \left[ \overline{W}^s(\sigma(\gamma(t), \sigma^1(t), ..., \sigma^z(t))) - \overline{W}(\sigma(\gamma(t), \sigma^1(t), ..., \sigma^z(t))) \right]$$
$$\text{for} \quad s = 1, ..., z - 1, \tag{a5}$$

where $\overline{W}^s(\sigma) = \sum_{i=1}^{u_s} \sigma_i^s W(P_{d_i^s}, \sigma(\gamma, \sigma^1, ..., \sigma^z))$ is the mean fitness in the $s$-th subpopulation. The argument of a fitness function is a set of relative frequencies of all individuals $\sigma$ (without division into subpopulations), therefore the opposite change of coordinates $\sigma(\gamma, \sigma^1, ..., \sigma^z)$ (a3) should be applied ([18]). In practical applications of this method to the modeling of biological problems, replicator equations can be defined for broken down populations. This break down will simplify the formulation of the model because, when strategies are initially assigned to subpopulations, there is no need to change their indices. The choice of subpopulations is arbitrary and depends on the biological assumptions underlying the analyzed problem. The entire



population may be divided into two competing subpopulations of carriers and parasites or predators and prey. It may also be divided into two subpopulations of males and females, in which case interspecific dynamics will describe the evolution of the secondary sex ratio, and intraspecific dynamics will describe changes of frequencies of strategies inside male and female subpopulations. The entire population can be divided into more than two subpopulations. The subpopulations can be divided into sub-subpopulations, and the entire population may be transformed into a complex multilevel cluster structure. However, all of these structures are equivalent to a single population replicator dynamics model.

## Appendix B

## Derivation of the fitness function of a gene

$$W_g(P_i, P, f, m) = W_g(P_i, G, M) = M_i W_m(P_i, P, f, m) + (1 - M_i) W_f(P_i, P, f, m) =$$

$$= \frac{P m_i}{G_i} W_m(P_i, P, f, m) + \frac{(1-P) f_i}{G_i} W_f(P_i, P, f, m) =$$

$$= \frac{k(1-P)}{2 G_i} \left( m_i \left( \sum_j f_j P_j + \frac{f_j}{m_j} P_i \right) + f_i \left( (1 - P_i) + \frac{m_i}{f_i} \left( 1 - \sum_j f_j P_j \right) \right) \right) =$$

$$= \frac{k(1-P)}{2 G_i} \left( m_i \sum_j f_j P_j + f_i P_i + f_i (1 - P_i) + m_i \left( 1 - \sum_j f_j P_j \right) \right) =$$

$$= \frac{k(1-P)}{2 G_i} \left( m_i \left( \sum_j f_j P_j + 1 - \sum_j f_j P_j \right) + f_j (P_i + 1 - P_i) \right) =$$

$$= \frac{k(1-P)}{2 G_i} (m_i + f_j)$$

The obtained formula should be described in new coordinates. Since:

$$m_i = \frac{M_i G_i}{P} \quad \text{and} \quad f_i = \frac{(1 - M_i) G_i}{1 - P},$$

in effect we obtain:

$$W_g(P_i, G, M) = \frac{k(1-P)}{2 G_i} \left( \frac{M_i G_i}{P} + \frac{(1 - M_i) G_i}{1 - P} \right) =$$



$$= \frac{k}{2}\left(M_i \frac{(1-P)}{P} + (1-M_i)\right).$$

## Appendix C

## Alternative formulation of the replicator dynamics

Derivation of replicator equations:

a) Dynamics of gene frequencies (6):

$$\dot{G}_i = G_i\left(W_g(P_i, P, f, m) - \overline{W}(P, f, m)\right) = G_i\left(\frac{k}{2}\left((1-M_i) + \frac{(1-P)}{P}M_i\right) - k(1-P)\right) =$$

$$= G_i k\left(\frac{(1-M_i)}{2} + (1-P)\left(\frac{M_i}{2P} - 1\right)\right) = G_i k\left(\frac{1}{2} - \frac{M_i}{2} + \frac{1-P}{P}\frac{M_i}{2} - 1 + P\right) =$$

$$= G_i k\left(\left(\frac{1}{P} - 2\right)\frac{M_i}{2} - \frac{1}{2} + P\right) = G_i k\left(\frac{1-2P}{P}\frac{M_i}{2} - \frac{1-2P}{2}\right) = G_i k\frac{1-2P}{2}\left(\frac{M_i}{P} - 1\right) =$$

$$= G_i k\left(\frac{1}{2} - P\right)\left(\frac{M_i}{P} - 1\right).$$

b) Dynamics of sex ratios in carriers subpopulations (7):

$$\dot{M}_i = M_i\left(W_m(P_i, P, f, m) - W_g(P_i, P, f, m)\right).$$

Since $\dfrac{f_i}{m_i} = \dfrac{(1-M_i)P}{M_i(1-P)}$ we have:

$$W_m(P_i, P, f, m) = k\frac{1-P}{2P}\left(\sum_j f_j P_j + \frac{f_i}{m_i}P_i\right) = k\frac{1-P}{2P}\left(\overline{P}_{pr} + \frac{(1-M_i)P}{M_i(1-P)}P_i\right).$$

Then equation $\dot{M}_i$ has the form:

$$\dot{M}_i = M_i\left(k\frac{1-P}{2P}\left(\overline{P}_{pr} + \frac{(1-M_i)P}{M_i(1-P)}P_i\right) - \frac{k}{2}\left((1-M_i) + \frac{(1-P)}{P}M_i\right)\right) =$$

$$= M_i\frac{k}{2}\left(\frac{1-P}{P}\left(\overline{P}_{pr} + \frac{(1-M_i)P}{M_i(1-P)}P_i\right) - (1-M_i) - \frac{(1-P)}{P}M_i\right) =$$

$$= M_i\frac{k}{2}\left(\frac{1-P}{P}\overline{P}_{pr} + \frac{(1-M_i)}{M_i}P_i - (1-M_i) - \frac{(1-P)}{P}M_i\right) =$$

$$= M_i\frac{k}{2}\left(\frac{1-P}{P}\left(\overline{P}_{pr} - M_i\right) + (1-M_i)\left(\frac{P_i}{M_i} - 1\right)\right) =$$



$$= \frac{k}{2}\left( M_i\left(\frac{1-P}{P}\right)\left(\overline{P}_{pr} - M_i\right) + \left(1 - M_i\right)\left(P_i - M_i\right)\right).$$

In effect, we obtain an alternative set of replicator equations (6) and (7).

## Appendix D

## Proof of Lemma 1

The equation of the sex ratio in the carrier subpopulations (7) can be denoted:

$$\dot{M}_i = \frac{k}{2}\left( M_i\left(\overline{P}_{pr} - M_i\right)\Gamma + \left(1 - M_i\right)\left(P_i - M_i\right)\right). \tag{d1}$$

At the stationary point, the right side of the equation should be equal to zero. The right side of this equation is a square polynomial of parameter $M_i$, then there exists at most two stationary points. Two terms are responsible for changing the direction of convergence: $\left(\overline{P}_{pr} - M_i\right)\Gamma$ and $\left(P_i - M_i\right)$ weighted by the current values of $M_i$ and $\left(1 - M_i\right)$. They are responsible for the attraction of $M_i$ suitably toward $\overline{P}_{pr}$ and $P_i$. If the current value of $M_i$ is smaller or larger than both values of $\overline{P}_{pr}$ and $P_i$, then both coefficients will have the same sign. If $\overline{P}_{pr} \neq P_i$, then both coefficients cannot attain zero in the same point. $M_i \in [0.1]$, and so it is obvious that the point that will zero the right side of equation should be contained in the interval limited by values of $\overline{P}_{pr}$ and $P_i$, because the terms will have opposite signs. It is also obvious that two stationary points cannot exist in the interior of the interval [0,1], because one should be an attractor and the second a repeller. This implies the existence of a third stationary point, which will be an attractor in the interval limited by a repeller and a boundary of the set [0,1]. Otherwise, the trajectory will escape the unit interval.

The interior has been analyzed. Thus we have to check the boundary of a set [0,1] where, the



second stationary point, a repeller, may exist. This may be 0, when $\overline{P}_{pr}$ or $P_i$ is equal to 0, or 1, when $\overline{P}_{pr}$ or $P_i$ is equal 1. Values of $\overline{P}_{pr}$ from a boundary of [0,1] are not biologically relevant [1], therefore we have to review two cases:

a) $P_i = 1$ and the possible restpoint $M_i = 1$.

b) $P_i = 0$ and the possible restpoint $M_i = 0$.

When we substitute $M_i = 1$ into a replicator equation $\dot{M}_i$, then $P_i$ vanishes, and the right side of equation (7) has a negative value, so this point is not stationary.

Thus, point a) is proven.

In the second case, when we substitute $M_i = 0$ to the equation (d1), we obtain:

$$\dot{M}_i = \frac{k}{2} P_i,$$

which means that for $P_i = 0$, there exists a stationary point in the boundary. Then, in general, for $P_i = 0$ equation (d1) takes the form:

$$\dot{M}_i = \frac{k}{2} M_i \left( \left( \overline{P}_{pr} - M_i \right) \Gamma - \left( 1 - M_i \right) \right). \tag{d2}$$

Therefore, there are two cases, $M_i = 0$ and $\left( \overline{P}_{pr} - M_i \right) \Gamma - \left( 1 - M_i \right) = 0$, for which the right side of the equation can go to zero. The second stationary point is $\overline{\overline{M}}_i = \dfrac{\overline{P}_{pr} \Gamma - 1}{\Gamma - 1}$. Bracketed term in (d2) is negative with respect to M_i only for $\Gamma > 1$, thus only in this case is $\overline{\overline{M}}_i$ stable. So we must check the following condition:

$$0 < \frac{\overline{P}_{pr} \Gamma - 1}{\Gamma - 1} \le 1. \tag{d3}$$

Thus $\overline{\overline{M}}_i \le 1$ for $\Gamma > 1$ when $\overline{P}_{pr} \le 1$ (relevant case) and for $\Gamma < 1$ when $\overline{P}_{pr} \ge 1$ (irrelevant case).



Thus, for the case $\Gamma > 1$, condition $\overline{\overline{M}}_i > 0$ should be checked. This leads to the condition

$$\overline{P}_{pr} > \frac{1}{\Gamma} \,.$$

After substitution of $\Gamma = \dfrac{1-P}{P}$ into obtained conditions we obtain:

$$\overline{P}_{pr} > \frac{P}{1-P} \quad \text{and} \quad P < \tfrac{1}{2}$$

When we parameterize $P = \dfrac{1}{a}$ where $a \in (1, \infty)$, we obtain:

$$\overline{P}_{pr} > \frac{\dfrac{1}{a}}{1 - \dfrac{1}{a}} = \frac{1}{a-1} > \frac{1}{a} \text{ (which means } \overline{P}_{pr} > P \text{)}$$

and

$$P < \tfrac{1}{2} \quad .$$

So this phenomenon is structurally stable, however, it exists only when $P < \dfrac{1}{2}$ and parameter $P$ is shifted from the current value of $\overline{P}_{pr}$. This means that it may be observed only at the beginning of convergence to the male subpopulation equilibrium (a rapid phase). Which is the proof of point b).

$\boxed{\cdot}$